\documentclass[pre,twocolumn,aps,showpacs,floatfix]{revtex4}

\usepackage{graphicx}
\usepackage{amsmath}
\usepackage{amsfonts}
\usepackage{amssymb}
\usepackage{epsf}
\usepackage{bm}

\begin{document}
\title{Observation of two-wave structure in strongly nonlinear
dissipative granular chains}
\author{Alexandre Rosas}
\affiliation{
Departamento de F\'{\i}sica,
Universidade Federal da Para\'{\i}ba,
Jo\~ao Pessoa, Para\'{\i}ba, Brazil, Caixa Postal 5008 - CEP: 58.059-970}
\author{Aldo H. Romero} 
\affiliation{
Cinvestav-Quer\'etaro,
Libramiento Norponiente 200, 
76230, Fracc. Real de Juriquilla,
Quer\'etaro, Quer\'etaro, M\'exico}
\author{Vitali F. Nesterenko}
\affiliation{
Department of Mechanical and Aerospace Engineering,
University of California San Diego,
La Jolla, CA 92093-0411}
\author{Katja Lindenberg}
\affiliation{
Department of Chemistry and Biochemistry,
and Institute for Nonlinear Science
University of California San Diego
La Jolla, CA 92093-0340}
\date{\today}

\begin{abstract}
In a strongly nonlinear viscous granular chain
under conditions of loading that exclude stationary waves
(e.g., impact by a single grain) 
we observe a pulse that consists of two interconnected but distinct
parts.  One is a leading narrow ``primary pulse" with properties
similar to a solitary wave in a ``sonic vacuum." It arises from strong
nonlinearity and discreteness in the absence
of dissipation, but now decays due to viscosity. 
The other is a broad, much more persistent shock-like ``secondary pulse"
trailing the primary pulse and caused by viscous dissipation.
The medium behind the primary pulse is transformed
from a ``sonic vacuum" to a medium with finite sound speed.
When the rapidly decaying primary pulse dies, the secondary pulse
continues to propagate in the ``sonic vacuum,"
with an oscillatory front if the viscosity is relatively small,
until its eventual (but very slow) disintegration.  
Beyond a critical viscosity there is no separation of the two pulses,
and the dissipation and nonlinearity dominate the shock-like
attenuating pulse
which now exhibits
a nonoscillatory front.
\end{abstract}

\pacs{46.40.Cd,43.25.+y,45.70.-n,05.65.+b}

\maketitle

Strongly nonlinear dissipative granular systems at conditions of loading
for which no stationary solution exists may exhibit new unexpected
behavior.
In this Letter we present the first observation of a new
phenomenon in a strongly nonlinear discrete viscous granular chain
subjected to a very short initial impulse.
The specific model under consideration is a chain of granules that
interact via a power law potential,
\begin{equation}
\begin{array}{l l l}
V(\delta_{k,k+1})&=\frac{A}{n}|\delta|^{n}_{k,k+1}, \qquad &\delta\leq
0,\\ \\
V(\delta_{k,k+1})&= 0, \qquad &\delta >0,
\end{array}
\label{eq:hertz}
\end{equation}
where $\delta_{k,k+1} \equiv y_k - y_{k+1}$ and $y_k$ is the
displacement of granule $k$ from its equilibrium position. 
In general the prefactor $A$ is a function of the Young's modulus $E$,
the Poisson ratio $\sigma$, and the principal radius of curvature $R$ of
the grain surfaces at the point of contact.  The exponent $n$ depends on
the topological properties of the contacting surfaces.
For the physically important case of the Hertz potential,
$n=5/2$ (spherical granules), $A =
[E/3(1-\sigma^2)]\sqrt{2R}$~\cite{landau}.
The equation of motion for the $k$th grain inside the chain is
\begin{eqnarray}
\ddot{x}_k &= &\left[\gamma \left(\dot{x}_{k+1} -\dot{x}_k\right) - (x_k
- x_{k+1})^{n-1}\right] \theta (x_k -
x_{k+1})  \nonumber\\
&&+\left[\gamma \left(\dot{x}_{k-1} -\dot{x}_k\right) + (x_{k-1} -
x_{k})^{n-1}\right] \theta (x_{k-1} - x_{k}),
\nonumber\\
\label{eq:motion}
\end{eqnarray}
where a dot denotes a derivative with respect to $t$, and $\gamma$ is
the viscosity coefficient. The Heaviside function $\theta(x)$ ensures
that the elastic and the viscous grain interactions exist only when the
grains are in contact.  Here we have introduced the rescaled variables
$x_k=y_k/b$, $t=(v_0/b)\tau$, $\gamma=\tilde{\gamma}(b/mv_0)$
[$b\equiv (mv_0^2/A)^{1/n}$] similar to those of Ref.~\cite{arosas}.
Note that the constant $A$ as well as the mass $m$ have been scaled out.
Initially the granules are placed side by side, just touching each other
but without precompression, and a velocity $v_0$ is imparted to a single
grain ($v_0=1$ in the scaled problem).

In the absence of the dissipative terms in Eq.~(\ref{eq:motion}), the
solution to the problem is well
understood as a result of extensive numerical and analytic
studies~\cite{arosas,nesterenko,nesterenkobook,hinch}.
For $n\geq 2$, analytic solutions in the long-wavelength approximation
agree with numerical simulations, and agreement is better for smaller
$n$.  One finds that the impact by a single particle (equivalent to a
$\delta$-function force applied to this particle) 
quickly develops into a stationary solitary wave whose
width/height depend on $n$.  For the physically
interesting case of elastic spherical grains ($n=5/2$) the
solitary wave resides on about five grains. 

The different approaches to the contact dissipation can be found
in~\cite{duvall,brilliantov,sen,falcon,brunhuber}.  Viscosity leads
to a number of salient effects.
In particular, dissipation based on relative velocities of granis
is able to dramatically change the pulse profile~\cite{herbold1,herbold2}.

In our case, below a critical viscosity, a pulse similar to a solitary wave 
caused by the strongly nonlinear
dispersive forces in the discrete medium is still generated. 
In subsequent discussion we shall call this pulse or its remnants
the \emph{primary pulse}.  However, because
the pulse is spatially narrow, there are high velocity gradients that
cause a relatively rapid loss of its energy. 
A quasistatic precompression appears behind the primary
pulse because the grains now
stop ``too early"~\cite{ourtwo}
as the pulse moves forward.  This precompression is
due entirely to dissipation and does not exist at zero viscosity.
It changes the nature of the medium
behind the primary pulse.  A broad \emph{secondary pulse} follows the
primary pulse.  This combination is an entirely
\emph{new structure}.  
The secondary pulse is appropriately thought of as a
``dissipative pulse" since it only occurs in the presence
of dissipation; it has much smaller velocity gradients than the primary
pulse and is therefore far more persistent.  It 
quickly evolves into a long-lived structure with a long 
tail of grains of uniform velocity.
In the following, we
discuss the detailed evolution of
these pulses, as well as the behavior of the excitation above the critical
viscosity.

\begin{figure}
\begin{center}
\includegraphics[width=8cm,height=6cm]{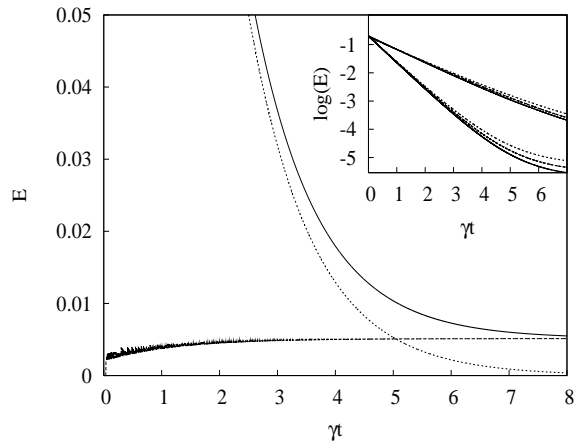}
\caption{Breakdown of the total energy (solid line) into portions
associated with the primary pulse (dotted) and the secondary pulse
(dashed) for $n=2.5$ and $\gamma=0.01$. Inset: time evolution of the
total energy, where two sets of three curves are shown, corresponding
to $n=2.2$ (upper set) and $2.5$ (lower set) in the Hertz potential.
Each set shows results for $\gamma=0.01$ (dotted), $0.005$ (dashed),
and $0.001$ (solid).
Early exponential decay reflects
the energy loss mainly by the primary pulse.  The energy
settles to a seemingly constant value, which is the remaining energy
stored in the secondary pulse.  This energy dissipates
on a much longer time scale than the initial loss.
\label{fig:breakdown}}
\end{center}
\end{figure}

Starting from the impact by a single grain corresponding to
an initial energy $E_0=1/2$, a small amount of energy is
lost in some back-scattering of nearby granules, but almost all of the energy
resides in the forward traveling wave, both parts of which are formed
very fast.  The total energy of the system as a function
of time is shown in the inset in Fig.~\ref{fig:breakdown} for
two values of $n$ and three values of $\gamma$.
The separation of the energy into the ``primary" and ``secondary"
portions for the case $n=2.5$ and $\gamma=0.01$ is shown in
Fig.~\ref{fig:breakdown}.  The primary pulse is a highly
nonstationary portion of the wave that
maximizes the rate of dissipation of some of the energy.  This is
reflected in the steep exponential decay associated with this loss. 
The energy decay slows down drastically as the primary pulse vanishes and
only the more persistent secondary pulse remains. Note that dissipation of
energy in the exponential decay regime is faster for
higher $n$. We explain this behavior by the larger velocity
gradients in the primary pulse, whose width decreases with increasing
$n$~\cite{nesterenkobook}.
An excellent numerical fit to the primary pulse
decay is provided by the expression $E(t)=E_0\exp(-0.92\gamma t)$,
while the rise of the secondary pulse is $A[1-\exp(-0.92\gamma t)]$
where the coefficient $A=0.005$ is the maximum energy of the secondary
pulse for these particular parameters.  As noted earlier, the
secondary pulse also decays eventually, but
far more slowly than the primary pulse, so that on
the time scale of Fig.~\ref{fig:breakdown}
the energy in the secondary pulse quickly reaches its maximum value and
remains essentially constant.  

The primary pulse travels along the chain with a diminishing speed
since its amplitude is decreased by the 
dissipation.  At very early times the secondary pulse also
exhibits a slightly diminishing amplitude and velocity, but increasing
energy, as it quickly
settles into a broad pulse of almost constant velocity amplitude with
a uniform velocity tail that stores kinetic energy.

\begin{figure}
\begin{center}
\includegraphics[width=9cm,height=6cm]{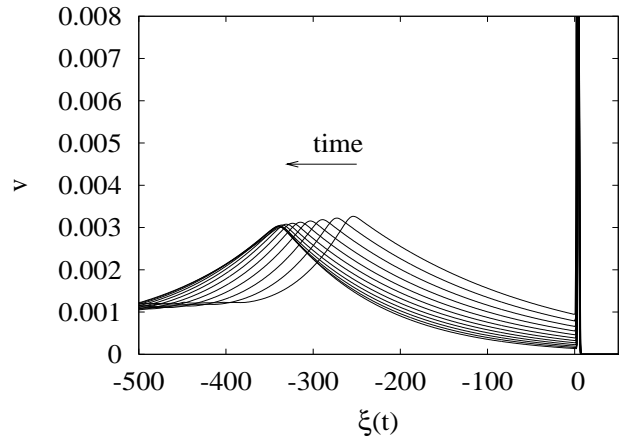}
\caption{
Snapshots of the velocities of the grains at early times ranging from
$40$ to $95$ in steps of $5$ in nondimensional units. The abscissa is the
moving variable $\xi(t)\equiv k-\int_0^t c(t)dt$, where $c(t)$ is the
time-dependent velocity of the primary pulse, and $k$ denotes the
granule in the chain.  
Eventually the primary pulse and the precompression behind it vanish
and the secondary pulse
continues to move at an essentially constant velocity in the sonic
vacuum.  Parameters are $n=5/2$ and $\gamma=0.005$.
\label{fig:secondary1}}
\end{center}
\end{figure}

\begin{figure}
\begin{center}
\includegraphics[width=9cm,height=6cm]{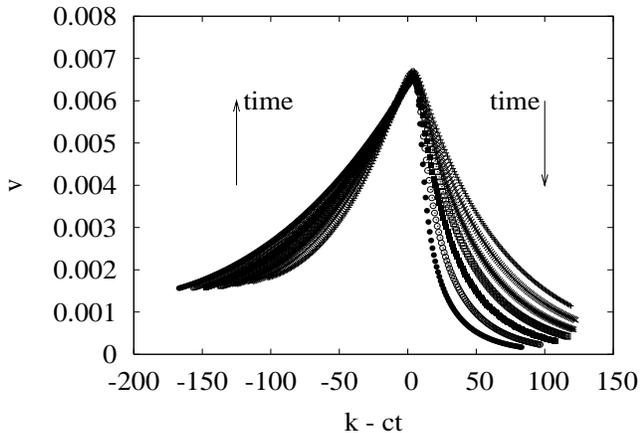}
\caption{
Velocity profile of the secondary pulse for $n=5/2$ and $\gamma=0.01$ at
different times, which increase upwards on the left side of the pulse
and downward on the right, consistent with the steepening of the
compression front of the pulse
with time.  The abscissa is now the moving variable with respect
to the essentially constant velocity of the peak of the secondary pulse.
\label{fig:profile}}
\end{center}
\end{figure}

Figures~\ref{fig:secondary1} and \ref{fig:profile} serve to detail the
behavior of the secondary pulse while the primary pulse has not yet
disappeared.  At first, the secondary pulse moves more slowly than the
primary, but this trend reverses as the primary pulse slows down with
its faster loss of energy and the peak of the secondary pulse acquires an
essentially constant velocity amplitude. Both figures show that the
secondary pulse is asymmetric, generating an extremely persistent tail
of essentially equal velocity granules behind it (not seen explicitly in
the figures).  
The practically constant velocity amplitude of the secondary pulse
is clearly seen
in Fig.~\ref{fig:profile}, where the abscissa for all the times shown is
scaled with the same constant velocity $c$.  It is 
interesting that this peak speed can be associated
with a local ``speed of sound" $c_s$, even though this system
is initially a sonic vacuum~\cite{nesterenko,nesterenkobook} and the
top portion of the pulse is only about $10$ particles wide.  
From the precompression $\Delta \equiv x_k - x_{k+1}$, where $k$ is the
position of the grain with maximum velocity, we obtain a dispersion
relation for the linearized chain whose coefficient is a local
``speed of sound" in the long wavelength approximation,
$c_s = \sqrt{\left(n-1\right) \Delta^{n-2}}$.
The value of $\Delta$ was obtained numerically. 
Table~\ref{table1} compares the
value of $c_s$ with that of the speed of the peak of the secondary
pulse for two potentials and several values of the damping parameter.
The agreement is clearly excellent.

\begin{table}
\begin{center}
\begin{tabular}{|c|c|c|c|c|c|}\hline
\multicolumn{3}{|c|}{$n=2.2$} & \multicolumn{3}{c|}{$n=2.5$} \\
 $\gamma$ & Pulse Vel. & $c_s$ &  $\gamma$ & Pulse Vel. & $c_s$ \\
\hline
 0.001 & 0.56 & 0.55 & 0.001 & 0.25 & 0.25\\ \hline
 0.005 & 0.66 & 0.66 & 0.005 & 0.38 & 0.38 \\ \hline
 0.01  & 0.71 & 0.71 & 0.01  & 0.44 & 0.44 \\ \hline
\end{tabular}
\end{center}
\caption{Comparison of peak pulse velocity and $c_s$ 
calculated from the precompression for two values of the potential
exponent $n$ and various values of the damping.}
\label{table1}
\end{table}

\begin{figure}
\begin{center}
\includegraphics[width=8cm,height=5.5cm]{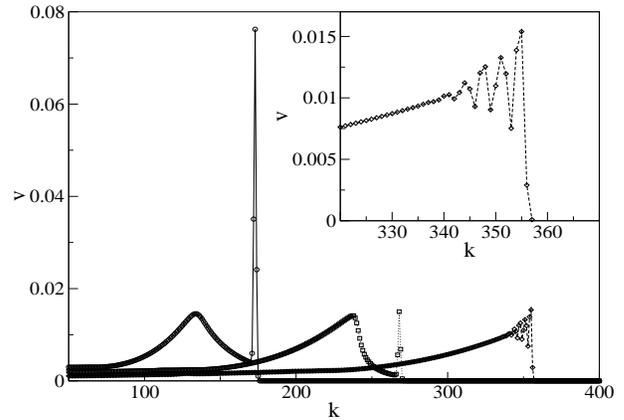}
\caption{
Snapshots of the velocity profile for small viscosity ($\gamma=0.02$) at
different times whose progression is easily recognizable as both pulses
move forward, the secondary pulse steepens, and the primary pulse
disappears.  The times are $500$, $900$, and $1400$ and $n=5/2$.
Inset: detailed view of the crest of the velocity profile at time
$1400$.
\label{fig:snap_g002}}
\end{center}
\end{figure}
 
Next we follow the continuing history of the pulses.
In Fig.~\ref{fig:snap_g002} we show a typical low-$\gamma$ unscaled 
progression with time.  The figure exhibits all the characteristics we have
discussed above, but it also shows three additional features.  One is
that the secondary pulse, being nonlinear, continues to change
in shape.  The pulse steepens (becoming more and more asymmetric) as
its peak travels faster (with local velocity $c_s$) than the
local sound speed at the bottom right of the peak. 
Secondly, the figure shows that the primary and secondary pulses have
\emph{comparable amplitudes} before the primary pulse dissipates.
Figure~\ref{fig:snap_g002} also shows that when the secondary pulse
is steep enough, dispersion starts to prevail and the
front displays oscillatory motion with structures that are a few grains
wide, similar to the primary pulse (see inset of Fig.~\ref{fig:snap_g002}).
The secondary pulse is shock-like, with velocities of the grains
in the pulse at least one order of magnitude smaller than the
pulse phase speed. 

The detailed results presented to this point are associated with small
values of $\gamma$.  In this regime it has been reasonable to speak of
two pulses as though they were separate entities, 
the primary being mainly due to nonlinearity and
discreteness, and the secondary one caused by dissipation and
nonlinearity.  The primary pulse causes the
precompression that underlies the secondary pulse, and in this sense
both together are a single entity.  Nevertheless, as a convenient manner
of speaking it is not inappropriate for these low viscosities to speak
of a ``separation" of pulses.

For \emph{small viscosities} ($\gamma \le 0.03$) the
secondary pulse reaches a critical slope \emph{before} catching the
primary pulse, while the primary pulse loses almost all of its
energy before being absorbed by the secondary pulse
(Fig.~\ref{fig:snap_g002}).  We have observed in computations
not presented here that in this small-$\gamma$ regime 
the maximum velocity in the secondary pulse increases
with increasing viscosity because larger dissipation is associated with a
greater precompression resulting in a secondary pulse of higher
amplitude.  We also point out that for very small
$\gamma$ ($\lesssim 0.002$) the secondary pulse has an almost
imperceptible amplitude on our numerical scale (and of course it
disappears entirely when $\gamma=0$), and the primary pulse has a
very long life. However we do not find a transition to a regime without
a secondary pulse for any finite value of $\gamma$.
The secondary pulse simply fades away smoothly with diminishing $\gamma$.

\begin{figure}
\begin{center}
\includegraphics[width=8cm,height=5.5cm]{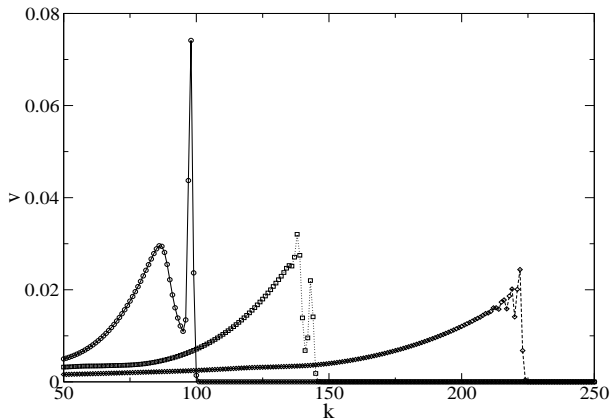}
\caption{
Snapshots of the velocity profile for intermediate viscosity
($\gamma=0.04$) at
different times: $140$, $220$,  and $400$, with $n=5/2$.
\label{fig:snap_g004}}
\end{center}
\end{figure}

For \emph{intermediate viscosities} ($0.04 \le \gamma \le 0.07$) the
secondary pulse catches up with the primary pulse while the primary
pulse still has energy comparable to the secondary (see
Fig.~\ref{fig:snap_g004}). As in the previous case, after the first pulse
disappears, the secondary pulse propagates as a shock-like wave with
an oscillatory front caused by the dispersion.

For \emph{large viscosities} ($\gamma \ge 0.07$) there
is no clear distinction between the primary and secondary pulses.

\begin{figure}
\vspace*{0.8cm}
\begin{center}
\includegraphics[width=8cm,height=5.5cm]{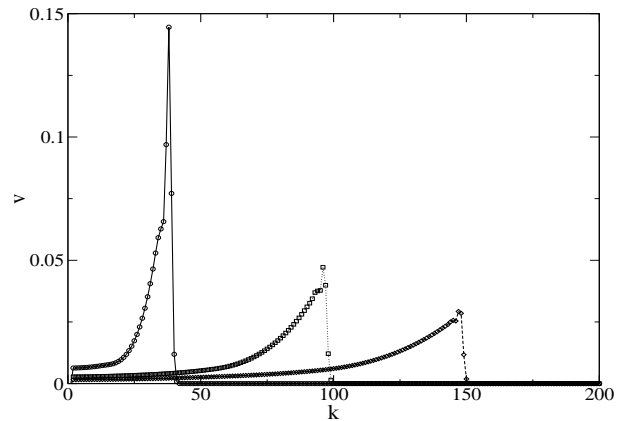}
\caption{
Snapshots of the velocity profile for large viscosity ($\gamma=0.1$) at
different times: $100$, $300$, and $500$, with $n=5/2$.
\label{fig:snap_g01}}
\end{center}
\end{figure}

For \emph{very large viscosities} ($\gamma \ge 0.1$) it is no longer
appropriate to think of two separate pulses
(Fig.~\ref{fig:snap_g01}).  From the beginning, there
is a single shock-like structure of dissipative origin with a sharp
essentially monotonic front. 

Recently Herbold and Nesterenko~\cite{herbold2}
found an oscillatory or monotonous shock-wave structure
(depending on viscosity) in a similar chain
but with different initial and boundary conditions in which the
velocity of the first particle is held constant (long input pulse).  

In summary, we have observed a new two-wave structure consisting of a 
dissipation-induced shock-like wave and a solitary-like
primary wave in a strongly nonlinear discrete granular chain
with no precompression. In our case the excitation is caused by
an impact by a single grain, equivalent to a
$\delta$-function force excitation.
In the presence of weak dissipation, the total pulse consists of two
qualitatively different parts.  A primary
pulse similar to that which is characteristic of a nondissipative sonic
vacuum is formed, but it is accompanied by a secondary pulse whose presence
is entirely due to viscosity.  Because of its high
velocity gradients, the primary pulse is rapidly attenuated, while the
broader and smoother secondary pulse persists for much longer times.  The
velocity of the maximum of the secondary pulse is practically constant
during its long lifetime, and its speed is essentially identical to
the local speed
of sound.  There are thus three distinctly separate time scales in
this problem: an extremely short scale for the formation of the double-pulse
excitation, a fairly rapid time scale of attenuation of the primary
pulse, and a very slow time scale for the eventual attenuation of the
secondary pulse.  Below a critical viscosity the secondary pulse develops
a dispersion-induced oscillatory front. 
Above the critical viscosity, it is no longer possible to
think of the primary and secondary pulses as separate entities, and the
resulting excitation presents a monotonic front.  
Similar behavior can be expected in other strongly nonlinear discrete
media under short pulse excitation.

This work was supported by the UC Institute for Mexico and the United
States (UC MEXUS) (AHR and KL), by 
the Conselho Nacional de Desenvolvimento Cient\'{\i}fico e Tecnol\'ogico
(CNPq) (AR), by CONACYT-Mexico project J-42647-F (AHR), and by
the National Science Foundation under grant No. PHY-0354937 (KL) and
grant No. DCMS03013220 (VFN).


\end{document}